\documentclass[aps,prl,twocolumn,superscriptaddress,preprintnumbers,amsmath,amssymb,floatfix,nofootinbib]{revtex4}
\usepackage{graphicx}
\usepackage{dcolumn}
\usepackage{bm}
\usepackage{url}
\usepackage{amsmath}
\usepackage{amssymb}
\usepackage{color}
\usepackage{subfig}
\usepackage{ulem}

\newcommand{\tbs}{$T_{bs}^-$ }
\newcommand{\bsud}{$bs\bar u\bar d$ }
\newcommand{\csud}{$cs\bar u\bar d$ }
\newcommand{\tcs}{$T_{cs}^0$ }

\begin{document}

\title{Weak-decay searches for $Qs\bar u\bar d$ tetraquarks}

\author{Fu-Sheng Yu}\email{yufsh@lzu.edu.cn}
\affiliation{School of Nuclear Science and Technology, Lanzhou University, Lanzhou 730000, China}
\affiliation{Frontiers Science Center for Rare Isotopes, and Lanzhou Center for Theoretical Physics, and Key Laboratory of Theoretical Physics of Gansu Province, Lanzhou University, Lanzhou 730000, China}
\affiliation{Center for High Energy Physics, Peking University, Beijing 100871, China}

\begin{abstract}

We propose to search for the open heavy-flavor exotic states with the quark components of $Qs\bar u\bar d$ with $Q=b,c$ via their weak decays. If there exist such exotic states below the $BK$ or $DK$ thresholds, they can only decay weakly. The advantages include: (i) the experimental backgrounds from the secondary decay vertex are much lower due to the long lifetimes, (ii) the productions are large enough for only one heavy quark in such states, (iii) the thresholds of $BK$ or $DK$ are about 260 MeV higher compared to the $B_s\pi$ or $D_s\pi$ thresholds of the configurations of $Qu\bar s\bar d$ or $Qd\bar s\bar u$ , with a higher possibility to exist such weakly-decay particles, and (iv) if observed, their inner structures are easier understood since they can not be induced by the kinematic effects.

\end{abstract}

\maketitle

The studies on exotic states beyond the conventional quark model are helpful to understand the non-perturbative properties of Quantum Chromodynamics (QCD). 
In the recent years, a lot of $XYZ$ and $P_c$ states have been observed,  explained by too many theoretical models, such as compact tetra- or penta-quarks, loosely bounded molecules, non-resonance by kinematic effects such as triangle singularities or cusp effects, and so on \cite{Chen:2016qju,Chen:2016spr,Olsen:2017bmm,Guo:2017jvc,Ali:2017jda,Esposito:2016noz,Lebed:2016hpi,Liu:2019zoy,Brambilla:2019esw,Guo:2019twa}. 
It always happens for the controversy in the theoretical understanding of the structures of the exotic states, due to the difficulty of non-perturbative QCD. 
It would be helpful if we find some new types of exotic states with less theoretical ambiguities. 

All of the observed exotic states are reconstructed via strong decays in experiment. 
In another word, they are all above some thresholds. 
Currently, there is no exotic state observed via a weak decaying process. 
If there exists a state below the corresponding lowest-lying flavor-conserving threshold, it can only decays weakly. 
It is therefore stable under the strong and electroweak interactions. 
The inner structure of a stable particle is relatively easier to be understood. With a long lifetime, it must be a bound state, but not just a peak induced by kinematic effects. 
If its mass was deeply below the threshold, it might be a candidate of compact tetraquark. 

The last discoveried weak-decay particle up to now is the double-charm baryon $\Xi_{cc}^{++}(ccu)$ by LHCb \cite{LHCb:2017iph}, via the decay mode of $\Lambda_c^+K^-\pi^+\pi^+$ suggested by a systematically study on the branching fractions \cite{Yu:2017zst}. The observation of $\Xi_{cc}^{++}$ implies a stable $bb\bar u\bar d$ tetraquark state \cite{Eichten:2017ffp,Karliner:2017qjm,Junnarkar:2018twb,bbud}. 
The mass of the doubly bottom tetraquark is predicted to be around 100 MeV below the $B^-\overline B^0\gamma$ threshold, leading to its decay only via the weak interaction. 
This is a good candidate of weak-decay tetraquark states. 
But unfortunately, it is very difficult to be observed currently due to the lower probability of production with two bottom quarks and the small branching fractions with both $b$-quark decays \cite{Ali:2018xfq,Ali:2018ifm}.
The open heavy-flavor tetraquarks are interesting for the weak decays, since the hidden heavy-flavor tetraquarks could at least annihilate the heavy flavor quark and its antiquark via strong or electromagnetic interactions.

The searches for open heavy-flavor exotic states have achieved a lot of progresses in experiments. 
$X_{0,1}(2900)$ with fully different quark flavor components of $cs\bar u\bar d$ was observed via a strong decay of $X_{0,1}(2900)\to D^+K^-$ in the process of $B^-\to D^-D^+K^-$ by LHCb \cite{LHCb:2020bls,LHCb:2020pxc}. 
Its explanations include most of the popular models like compact tetraquarks, molecules, triangle singularities and so on \cite{Liu:2020nil,Liu:2020orv,Chen:2020aos,Chen:2020eyu,csud}. 
Very recently, LHCb observed the first double-charm tetraquark $T_{cc}^+(cc\bar u\bar d)$ \cite{LHCb:2021vvq,LHCb:2021auc}. 
Its mass is lower than and very close to the threshold of $D^0D^{*+}$. But it still decays strongly into $D^0D^{0}\pi^+$. 
Its production and decay have been studied in \cite{Qin:2020zlg} with the predicted signal yields manifested by the LHCb measurement, supporting a picture of compact tetraquark \cite{Jin:2021cxj}.
The molecular state is also widely used to explain its inner structure \cite{Li:2021zbw,Dong:2021bvy,Xin:2021wcr,ccud}. 
Another interesting candidate is $X(5568)$ which was found in the final state $B_s^0\pi^\pm$ by the D0 collaboration \cite{D0:2016mwd}, with four different quark flavors of $bd\bar u\bar s$ or $bu\bar d\bar s$. 
But subsequently this result was not confirmed by other experiments \cite{Aaij:2016iev,CMS:2017hfy,CDF:2017dwr,ATLAS:2018udc}. 
All of the above  states are measured via their strong decays. 
On the other hand, the observations or evidences of them implies that the productions of such open heavy-flavor exotic states are large enough.

In this work, we propose to search for the four-quark states with the quark components of $Qs\bar q_1\bar q_2$, with $Q=b,c$ and $q_{1,2}=u,d$, in the case that their masses are below the $BK$ or $DK$ thresholds so that they can only decay weakly. 
There are several advantages to search for such weak-decay open heavy-flavor exotic states. 
Firstly, the experimental backgrounds from the secondary decay vertex are much lower due to the long lifetimes. 
Secondly, the productions are large enough for only one heavy quark in such states. 
Thirdly, the thresholds of $BK$ or $DK$ are about 260 MeV higher compared to the $B_s\pi$ or $D_s\pi$ thresholds of the configurations of $Qq_1\bar s\bar q_2$, with a higher possibility to exist such weak-decay particles. 
The large difference between the two thresholds results from the relatively larger masses of kaons compared to those of pions, which violate the $SU(3)$ chiral symmetry. 
Finally, if observed, their inner structures are easier understood since they must be bound states and can not be induced by the kinematic effects.

In order to search for such weak-decay exotic particles, the key issues include their masses, productions, lifetimes and branching fractions of some favored decaying processes. 
We will discuss them in details in the following. 
Since we are discussing on searching for possibly weak-decay particles, only the lightest states with the flavor components of \bsud and \csud are considered in this paper, denoted as \tbs and \tcs, respectively.

\begin{figure}[!]
\includegraphics[scale=0.13]{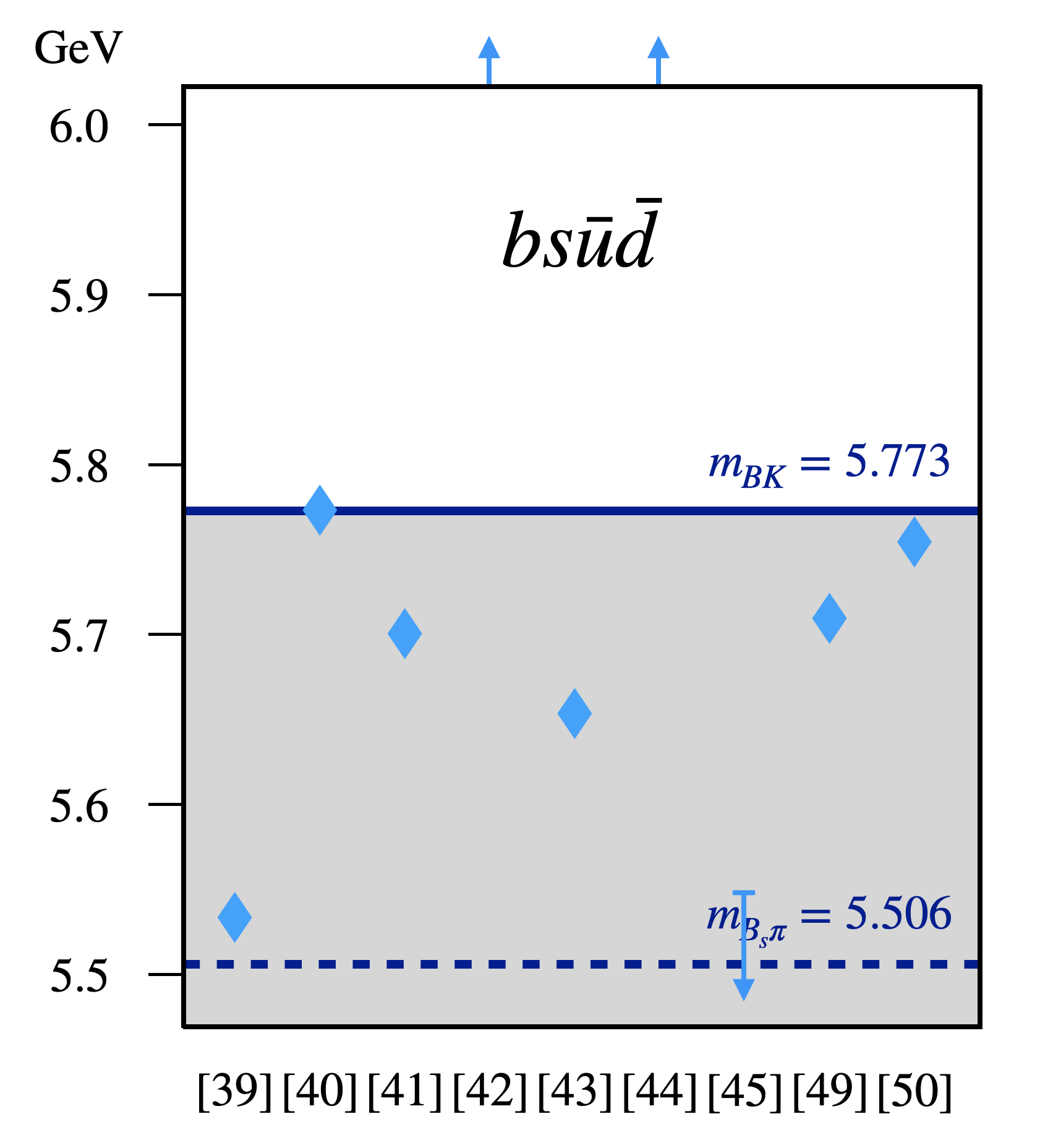}
\includegraphics[scale=0.13]{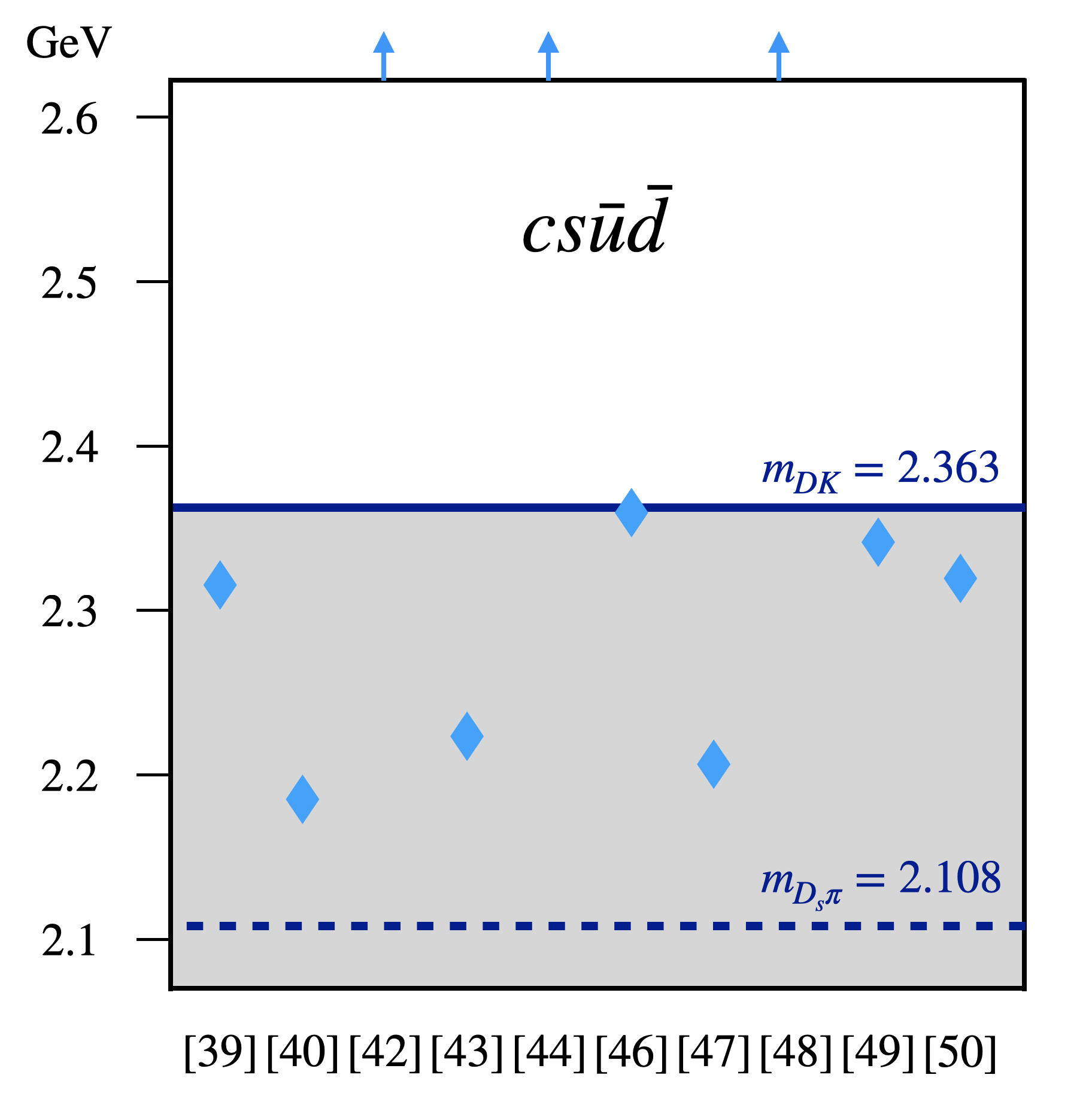}
\caption{Masses of the lightest states of $bs\bar u\bar d$ and $cs\bar u\bar d$ tetraquarks predicted in literature.}\label{fig:mass}
\end{figure}

The masses of \tbs and \tcs have been widely investigated in the literature motivated by $X(5568)$ and $X_{0,1}(2900)$. 
We show the results given by systematical studies on the corresponding spectrum \cite{Liu:2016ogz,He:2020jna,Guo:2021mja,He:2016xvd,Cheng:2020nho,Chen:2018hts,Yang:2021izl,Huang:2019otd,Xue:2020vtq,Kong:2021ohg,Sundu:2019feu,Wang:2020prk} in Fig. \ref{fig:mass}. 
It can be seen that several theoretical works have predicted the existence of \tbs and \tcs below the $BK$ or $DK$ thresholds, while some predictions are above the thresholds.
All these predictions show that \tbs and \tcs are the state of $J^P=0^+$ with the $\bar u\bar d$ diquark being $0^+$ and color ${\bf \bar 3}$.
No such bound states are present in some other analysis \cite{Hudspith:2020tdf,Cheung:2020mql}. 
Due to the difficulty of non-perturbative QCD calculations, we will not make a conclusion on the mass predictions. 
The purpose of this work is to propose a good method to search for them via weak decays if there exist the strong-interaction stable states. 

In the case that \tbs exists below the $BK$ threshold, its lifetime can be easily estimated to be around 1.5 ps. 
There is only one heavy quark, the bottom quark, while all the other quarks are light. 
Therefore, its lifetime is dominated by the $b$ quark decays. 
In the heavy quark limit, the lifetimes of all the charmless $b$-flavored hadrons are close to each other \cite{PDG}, 
\begin{align}
\tau(T_{bs}^-)&\approx \tau(B^-)\approx \tau(\overline B^0)\approx \tau(\overline B_s^0)
\nonumber\\
&\approx \tau(\Lambda_b^0)\approx \tau(\Xi_b^-)\approx \tau(\Xi_b^0)\approx1.5 \,\text{ps}.
\end{align}
The non-spectator effect from the $s$ quark is similar to those of $\Xi_b^{-,0}$, while the effects from $\bar u$ and $\bar d$ anti-quarks are in analogy with those of $B^{+,0}$. 
All these effects are corrections at the order of $1/m_b^3$ in the heavy quark expansion, and are expected to be around $10\%$ or less \cite{Lenz:2014jha}.
The $b\bar d\leftrightarrow d\bar b$ oscillation might happen in $T_{bs}^-$. 
But after oscillating to $[ds][\bar b\bar u]$, it would immediately strongly decay into $B_{s}^{0}\pi^{-}$. 
Its contribution to the lifetime of \tbs is negligible since it happens via box diagrams compared to the tree contributions. 

\begin{figure}[!]
\includegraphics[scale=0.17]{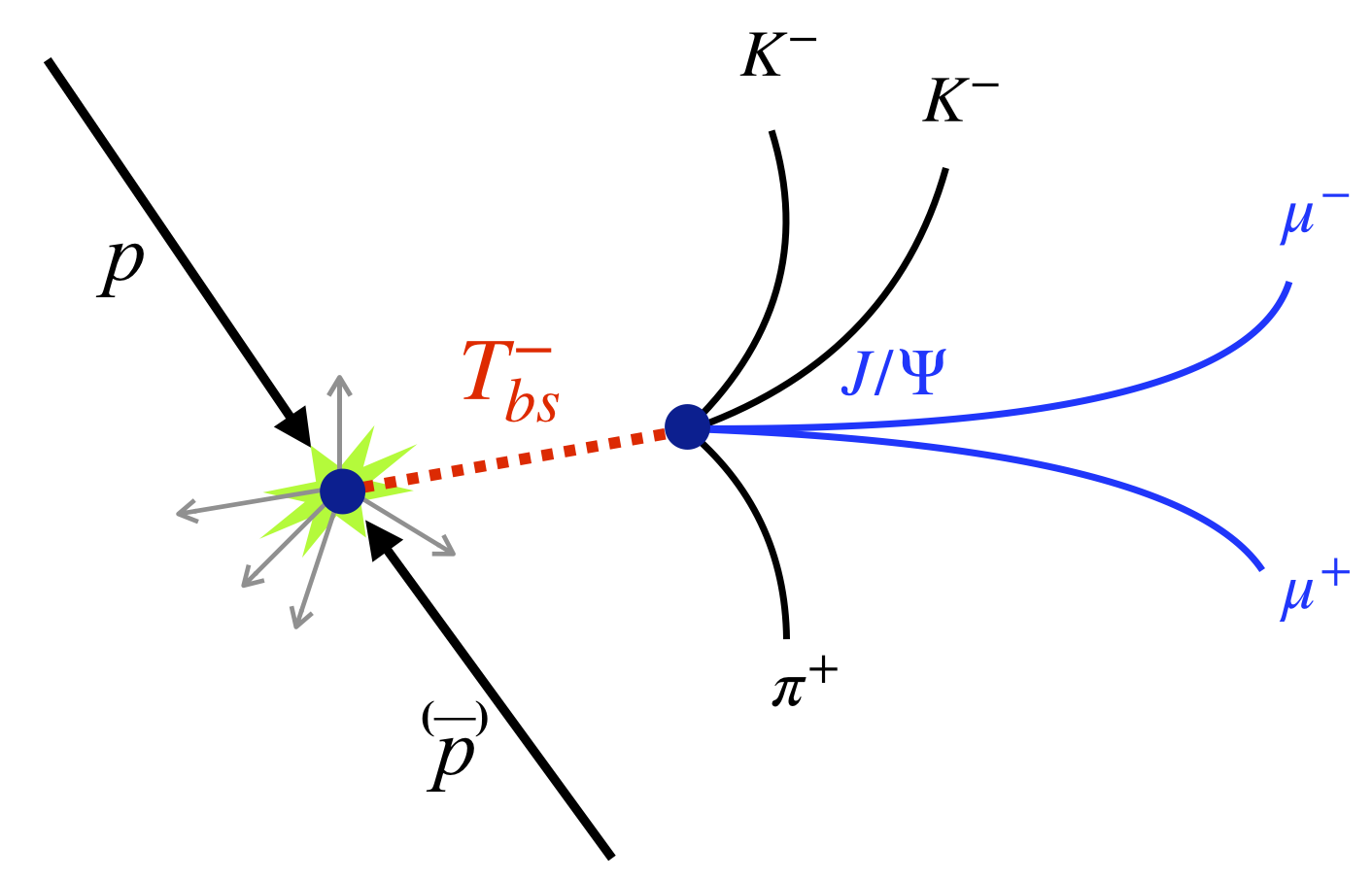}
\includegraphics[scale=0.17]{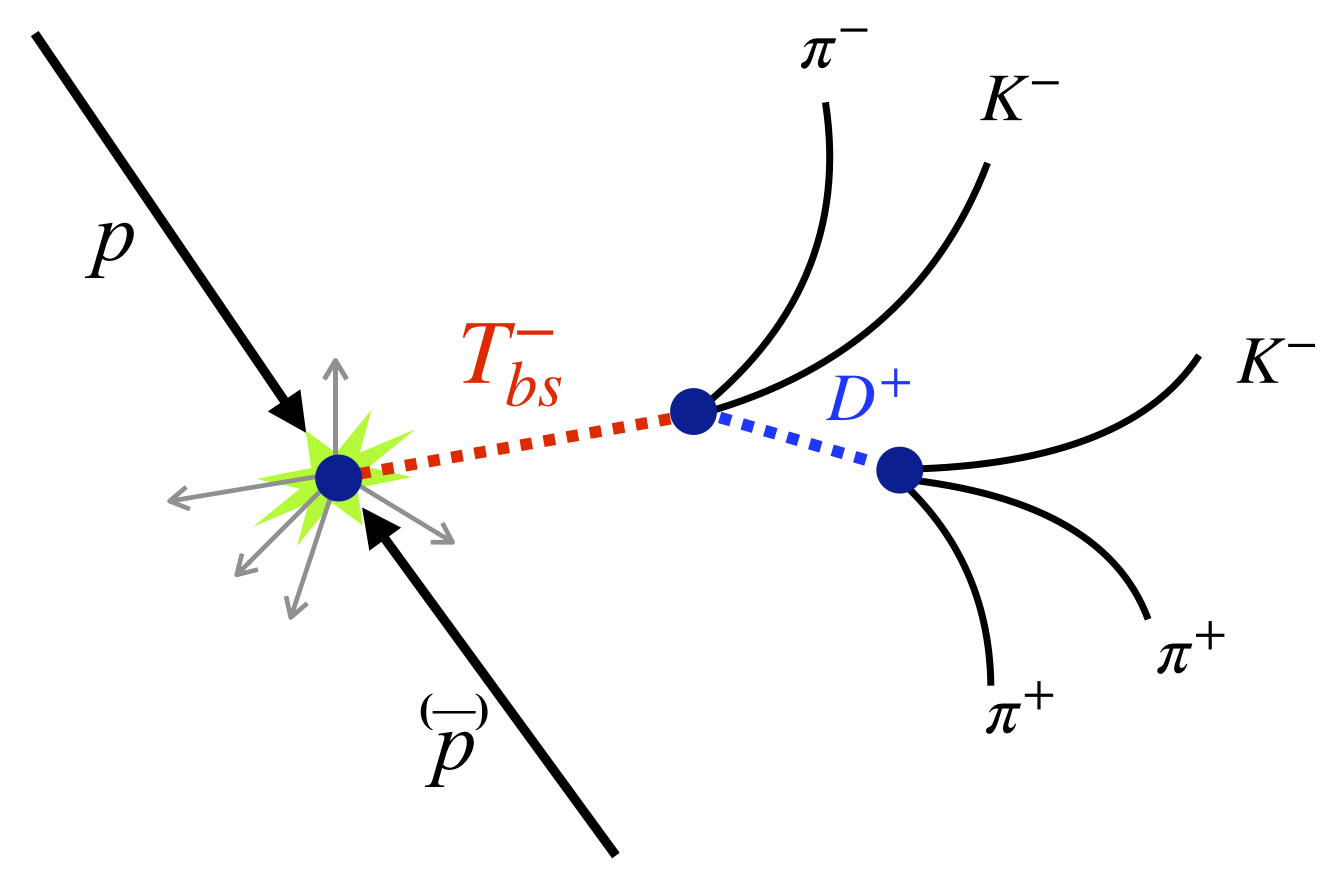}
\caption{Depicted diagrams of the weak decays of the below-threshold \bsud tetraquark.}\label{fig:topo}
\end{figure}

One of the key points of this work is that a particle with a longer lifetime is easier to be observed in the hadron colliders \cite{Yu:2019lxw}. 
Similarly to ordinary ground-state $B$ mesons and baryons, the weak-decay \tbs is boosted to fly a distance in the detector after its production in the high-energy hadron colliders. 
The final-state tracks come from a common vertex where \tbs decays, which is displaced from the primary $p\bar p$ (or $pp$) collision vertex where \tbs is produced. 
The lifetime of \tbs is long enough to distinguish the primary vertex and secondary vertex at Tevatron and LHC, like the ordinary $B$ mesons and $b$-baryons. 
Therefore, the stable \tbs benefits a much lower background, making it much easier to be observed.

\begin{figure}[!]
\includegraphics[scale=0.37]{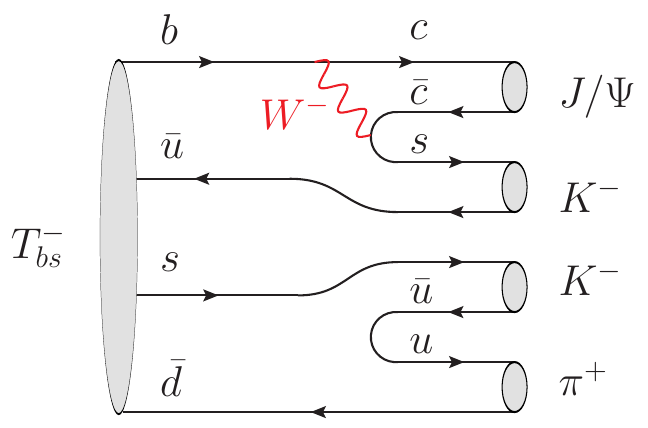}
\includegraphics[scale=0.14]{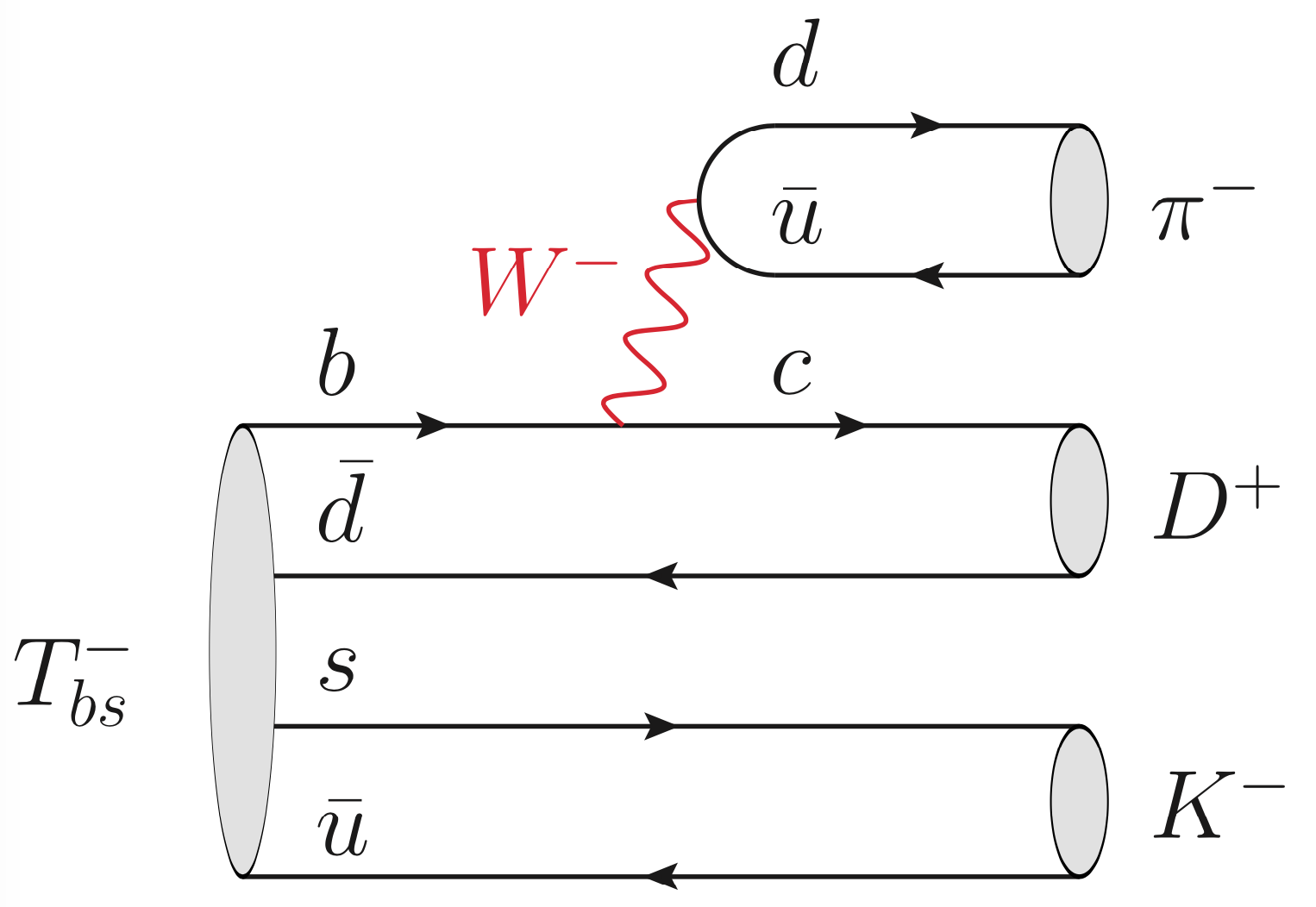}
\caption{Topological diagrams of weak decays of the \bsud tetraquark into $J/\Psi K^{-}K^{-}\pi^{+}$ and $D^+K^-\pi^-$, respectively.}\label{fig:feyn-bsud}
\end{figure}

The most favorable weakly decay modes are \tbs$\to J/\Psi K^{-}K^{-}\pi^{+}$ and $D^+K^-\pi^-$. 
All the final-state particles are electrically charged so that can be easily detected in the hadron colliders, seen in Fig. \ref{fig:topo} for the depicted diagrams in the collisions.
The dominant topological decaying diagrams are shown in Fig. \ref{fig:feyn-bsud}. 
The decay of \tbs$\to J/\Psi K^{-}K^{-}\pi^{+}$ is dominated by the color-suppressed emitted tree diagram, while \tbs$\to D^+K^-\pi^-$ by the color-favored emitted tree diagram. 
Both of them contribute to relatively large branching fractions \cite{Xing:2019hjg}. 
The branching fractions are approximately similar to those of corresponding $B$ meson decays at the leading order of the heavy quark expansion, 
\begin{align}\label{eq:Br1}
\mathcal{B}(T_{bs}^-\to& J/\Psi K^{-}K^{-}\pi^{+})
\approx \mathcal{B}(B^{-}\to J/\Psi K^{-})\nonumber\\
&\approx \mathcal{B}(\overline B^{0}\to J/\Psi K^{-}\pi^+)
=\mathcal{O}(10^{-3}),\\
\mathcal{B}(T_{bs}^-\to& D^+K^-\pi^-)
\approx\mathcal{B}(B^{-}\to D^0\pi^{-})\nonumber\\
&\approx\mathcal{B}(\overline B^{0}\to D^+\pi^{-})
=\mathcal{O}(10^{-3}).
\end{align}

The prompt production of \tbs at the hadron colliders can be directly compared to the known productions of $B$ mesons and $b$-baryons. The production of heavy-light hadrons are usually understood as the production of a heavy-flavor quark and the fragmentation into the corresponding hadrons. 
The ratio of cross sections can be expressed by the fragmentation functions as in the following
\begin{align}
{\sigma(pp\to T_{bs}^-X)\over\sigma(pp\to B^-X)}={f_{T_{bs}}\over f_{B}}={f_{T_{bs}}\over f_{\Xi_b}}{f_{\Xi_b}\over f_{\Lambda_b}}{f_{\Lambda_b}\over f_{B}},
\end{align}
where $f_i$'s are the fragmentation functions of a $b$ quark into the corresponding hadrons. Here, $f_B$ is exactly the same as $f_u$ or $f_d$ which are usually used in some literature.
The ratios of the right-handed side of the above equation can be obtained from the current measurements. 
The productions of $T_{bs}^-$($bs\bar u\bar d$) relative to $\Xi_b(bsu)$ can be taken approximately as the fragmentations with a $\bar u\bar d$ diquark v.s. a $u$ quark, $f_{T_{bs}}/f_{\Xi_b}\propto f_{\Lambda_b}/f_B$.
Note that the excited states of \bsud are probably above the $BK$ thresholds, thus would decay strongly into $BK$ or other final states, but not $T_{bs}^-$.
This is different from $B$ mesons and $b$-baryons whose excited states will definitely decays into the ground states of the bottom hadrons. 
The fragmentation of $b\to T_{bs}^-$ should thus be taken as the primary fragmentation functions of the ground states, just like the case of $T_{cc}^+(cc\bar u\bar d)$ as done in \cite{Qin:2020zlg}. 
The primary fragmentation functions, represented as $f_i'$, are according to the direct production of hadrons from high-energy collisions, without the feeddown from the excited states.
Therefore, it can be obtained that  ${f_{T_{bs}}/f_{\Xi_b}}\sim{f_{\Lambda_b}'/f_B}=(f_{\Lambda_b}'/f_{\Lambda_b})(f_{\Lambda_b}/f_B)$.
The primary fragmentation of ground state of $\Lambda_c$ has been measured by Belle that $f_{\Lambda_c}'/f_{\Lambda_c}=0.48\pm0.08$ \cite{Belle:2017caf}.
Under the heavy quark symmetry, the ratio of $f_{\Lambda_b}'/f_{\Lambda_b}$ would be close to that of the charmed baryons. 
The baryon-to-meson ratio of fragmentations has been measured by LHCb with $f_{\Lambda_b}/f_B=0.518\pm0.036$ at the averaging range of $4<p_T<25$GeV and $2<\eta<5$ \cite{LHCb:2019fns}. 
Currently, ${f_{\Xi_b}/ f_{\Lambda_b}}$ has a relatively large uncertainty. 
It is measured that ${f_{\Xi_b}/ f_{\Lambda_b}}=0.082\pm0.027$ by LHCb via $\Xi_b^-\to J/\Psi\Xi^-$ and $\Lambda_b^0\to J/\Psi\Lambda$ under the assumption of the flavor $SU(3)$ symmetry \cite{LHCb:2019sxa}, which is consistent with the theoretical prediction given by \cite{Jiang:2018iqa}. 
The ratio of productions of \tbs compared to $B$ meson is finally obtained as $0.011\pm0.004$.

With all the information of productions and decays, we can estimate the signal yields of $T_{bs}^-$. 
We compare the decays of \tbs$\to J/\Psi K^-K^-\pi^+$ and $D^+K^-\pi^-$ with the $B$ meson decaying processes with the same number of tracks, in order to control the detection efficiencies of final particles, such as $B^-\to J/\Psi K^-\pi^+\pi^-$ and $B^-\to D^+\pi^-\pi^-$. 
Their branching fractions are $(0.81\pm0.13)\times10^{-3}$ and $(1.07\pm0.05)\times 10^{-3}$ \cite{PDG}, respectively.
In \cite{LHCb:2020fvo}, $5.5\times10^5$ signal events of $B^-\to J/\Psi K^-\pi^+\pi^-$ were found using 9 fb$^{-1}$ data at LHCb. 
 It was reported that $4.9\times10^4$ events of $B^-\to D^+\pi^-\pi^-$ were collected for the 
3 fb$^{-1}$ data by LHCb \cite{LHCb:2015eqv}.
Considering the similar lifetimes and decaying branching fractions between \tbs and $B^-$, and the ratio of productions, it can be expected that the signal yields of  \tbs$\to J/\Psi K^-K^-\pi^+$ and $D^+K^-\pi^-$ would be at the order of $10^{2}$ or $10^{3}$ with the current LHCb data. 
With the low backgrounds due to the long lifetime, such large number of events is a good opportunity for the discovery, as long as there exists a weak-decay $T_{bs}^-$.
If \tbs were not observed via the above two decay processes, it would help to exclude the existence of the tetraquark state of \bsud  below the $BK$ threshold. 

The search strategy of \tcs is slightly different from that of \tbs, since the heavy quark expansion in the charmed hadrons does not work as well as that in the bottom hadrons.
If \tcs exists below the $DK$ threshold, it could weakly decays into $K^{-}K^{-}\pi^{+}\pi^{+}$ via the transition of $c\to us\bar d$. 
The branching fractions of Cabibbo-favored processes of charmed hadron decays are usually at the order of percent. 
Unlike the strategies for $T_{bs}^-$, prompt productions are not suitable for the searches of $T_{cs}^0$, although the productions of charmed hadrons are even larger than the bottom ones.
The lifetime of \tcs would be shorter than \tbs by around one order of magnitude. The heavy quark expansion does not work well in the charm system, so then the spectators contribute significantly to the total width of the charmed hadrons. For example, the lifetimes of charmed baryons are ranging from 0.15 ps to 0.45 ps, smaller than those of charmed mesons from 0.41 ps to 1.04 ps, and much smaller than those of bottom hadrons with 1.5 ps. So then the lifetime of \tcs are expected to be smaller than charmed baryons with even one more spectator quark. 
Such short lifetime would make it difficult to distinguish the secondary vertex of \tcs decays from the primary vertex of prompt productions, so that difficult to lower down the backgrounds. 
Beside, the combined backgrounds of $K^{-}K^{-}\pi^{+}\pi^{+}$ are too large at the proton-proton collision. 
Therefore, it should be changed for the strategy to search for \tcs via $B$ meson decays.

\begin{figure}[!]
\includegraphics[scale=0.37]{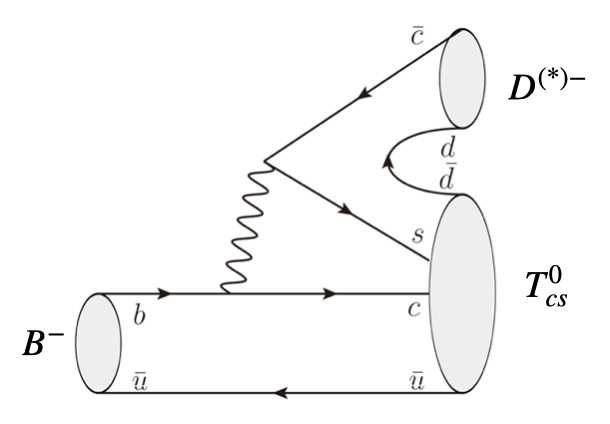}
\includegraphics[scale=0.37]{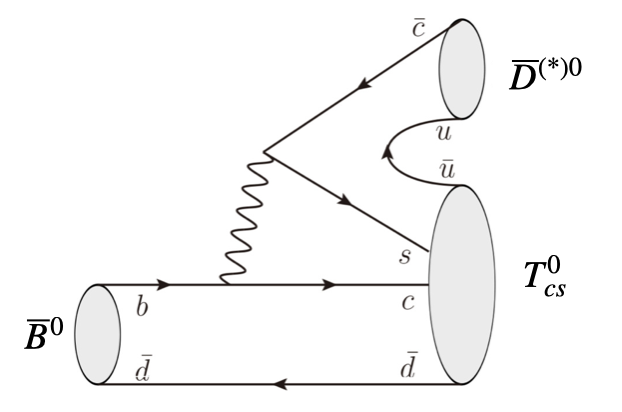}
\caption{Topological diagrams of $B$ mesons decaying into $T_{cs}^0$.}\label{fig:feyn-csud}
\end{figure}

We propose to measure $B^-\to D^{(*)-}$\tcs and $\overline B^0\to \overline D^{(*)0}$\tcs with \tcs$\to$$K^{-}K^{-}\pi^{+}\pi^{+}$, whose topological diagrams are shown in Fig.\ref{fig:feyn-csud}. 
In this case, it could avoid the large backgrounds. 
Besides, the observation of $B\to DX_{0,1}(2900)$ implies that the production of \tcs in $B$ decays might be large enough as well. 
The quark flavors of the final states have two strange quarks and two down anti-quarks, which could not form any ordinary $q\bar q$ mesons.
Even if a mis-identification between a pion and a kaon in experiments, the width of the weak-decay \tcs is extremely narrow, compared to the relatively large width of a resonance with a mass above 2 GeV. 

In conclusion, we propose to search for the four different flavor exotic tetraquark states of \bsud and \csud via their weak decays, if there exist such states below the $BK$ or $DK$ thresholds. The advantages include larger productions, longer lifetimes lowering down the experimental backgrounds, higher thresholds and more clear inner structures. 
The searching for the \bsud tetraquark could be performed by measuring the processes of \tbs$\to J/\Psi K^-K^-\pi^+$ and $D^+K^-\pi^-$. 
The \csud tetraquark could be searched for via the decays of $B^-\to D^{(*)-}$\tcs and $\overline B^0\to \overline D^{(*)0}$\tcs with \tcs$\to$$K^{-}K^{-}\pi^{+}\pi^{+}$.
They could be easily measured using the current LHCb data. 
If not observed, it would also be helpful to constrain the lower limit of the masses of such tetraquarks.

\section{Acknowledgement}
We are grateful to Hai-Yang Cheng, Shi-Yuan Li, Xiang Liu, Yan-Rui Liu, Cai-Dian L\"u and Wei Wang for the theoretical discussions. We are also grateful to Yuan-Ning Gao, Zhen-Wei Yang, Li-Ming Zhang and Yan-Xi Zhang for the discussions on the experimental measurements. 
This work was supported in part by the National Natural Science
Foundation of China under Grants No. 11975112.

%%%%%%%%%%%%%%%%%%%%%%%%%%%%%%%%%%%%%%%%%%%%%%%%%%%%%%%%%%%%%%%%%%%%%%%%%%%%%%%%%%%%%%%%%%%%%%%%%%%%%%%%%%%%%%%%%%%%%%%%%%%%

\end{document}